\documentclass[a4paper,twocolumn,accepted=2023-09-06,citeautoscript,pra,superscriptaddress,amsmath,amssymb,longbibliography]{quantumarticle}
\pdfoutput=1

\usepackage{mathtools}
\usepackage{ifxetex,ifluatex}
\usepackage{graphicx}
\usepackage{float}
\usepackage[unicode=true,
  colorlinks=true,
  linkcolor=quantumviolet,
  citecolor=quantumviolet]{hyperref}
\hypersetup{breaklinks=true, pdfborder={0 0 0}}
\usepackage{bm}
\usepackage{braket}
\usepackage{xcolor}

\begin{document}
\title {Noise-robust ground state energy estimates from deep quantum circuits}

\author{Harish~J.~Vallury}
\affiliation{School of Physics, University of Melbourne, Parkville, VIC 3010, Australia}
\orcid{0009-0007-3355-5458}

\author{Michael~A.~Jones}
\affiliation{School of Physics, University of Melbourne, Parkville, VIC 3010, Australia}
\orcid{0000-0002-4060-3493}

\author{Gregory~A.~L.~White}
\affiliation{School of Physics, University of Melbourne, Parkville, VIC 3010, Australia}
\orcid{0000-0001-6673-6676}

\author{Floyd~M.~Creevey}
\affiliation{School of Physics, University of Melbourne, Parkville, VIC 3010, Australia}
\orcid{0009-0008-9088-5266}

\author{Charles~D.~Hill}
\affiliation{School of Physics, University of Melbourne, Parkville, VIC 3010, Australia}
\affiliation{School of Mathematics and Statistics, University of Melbourne, Parkville, VIC 3010, Australia}
\orcid{0000-0003-0185-8028}

\author{Lloyd~C.~L.~Hollenberg}
\affiliation{School of Physics, University of Melbourne, Parkville, VIC 3010, Australia}
\orcid{0000-0001-7672-6965}

\begin{abstract}
{
In the lead up to fault tolerance, the utility of quantum computing will be determined by how adequately the effects of noise can be circumvented in quantum algorithms. Hybrid quantum-classical algorithms such as the variational quantum eigensolver (VQE) have been designed for the short-term regime. However, as problems scale, VQE results are generally scrambled by noise on present-day hardware. While error mitigation techniques alleviate these issues to some extent, there is a pressing need to develop algorithmic approaches with higher robustness to noise. Here, we explore the robustness properties of the recently introduced quantum computed moments (QCM) approach to ground state energy problems, and show through an analytic example how the underlying energy estimate explicitly filters out incoherent noise. Motivated by this observation, we implement QCM for a model of quantum magnetism on IBM Quantum hardware to examine the noise-filtering effect with increasing circuit depth. We find that QCM maintains a remarkably high degree of error robustness where VQE completely fails. On instances of the quantum magnetism model up to 20 qubits for ultra-deep trial state circuits of up to $\sim$500 CNOTs, QCM is still able to extract reasonable energy estimates. The observation is bolstered by an extensive set of experimental results. To match these results, VQE would need hardware improvement by some 2 orders of magnitude on error rates.
}
\end{abstract}

\maketitle

\section{Introduction}

As increasingly complex quantum devices emerge from current fabrication capabilities~\cite{ebadi2021quantum,mi2021information,mooney2021whole,frey2022realization}, a major question we face is the extent to which this technology can transition from fascinating physics experiments into useful information processing~\cite{montanaro2016quantum,shor1994algorithms,gidney2021factor,aspuru-guzik}. An open problem in the noisy intermediate-scale quantum (NISQ) regime~\cite{nisq} is: on which side of the fence can noisy computations fall, classically simulable or quantumly useful? 
Thus far, the rapid development of the field in the past few years has resulted in the availability of fully programmable NISQ devices on the scale of hundreds of qubits~\cite{gambetta2020ibm,morgado2021quantum}, and certain purpose-built algorithms run on these devices have demonstrated the ``quantum supremacy" milestone by outperforming their classical counterparts~\cite{supremacy,supremacy2}. A major challenge for the field now is to obtain some form of practical quantum advantage for real-world problems on NISQ hardware \cite{daley2022practical}. A popular school of thought to this effect is to create complicated quantum states and then extract meaningful observables~\cite{georgescu2014quantum,kandala,cao2019quantum}. 
One axis for exploration with respect to the noise problem is the choice of measured properties, and how these might either be hindered by or circumvent the accumulation of quantum errors. In particular, here we study higher weight corrections to ground state energy estimates and the interplay of these observables with real-world noise.

A common hybrid approach for solving problems on NISQ is the variational quantum eigensolver (VQE)~\cite{vqe}, which determines estimates for the ground state of a given quantum system, with practical application for chemistry and materials problems~\cite{vqe,kandala}. The subfield of variational quantum computing has been burgeoning over the recent years~\cite{fedorov2022vqe} -- with developments in the efficiency and scope of state preparation via adaptive algorithms~\cite{grimsley2019adaptive,tang2021qubit} and symmetry preservation~\cite{gard2020efficient,seki2020,anselmetti2021local}, and measurement of observables~\cite{santagati2018witnessing,hamamura2020efficient,huang2021efficient}. However, despite favourable NISQ-era prospects and indications that stochasticity can assist in optimisation~\cite{liu2022noise}, the general viability of hybrid variational approaches on present-day hardware is still heavily impacted by the overwhelming effect of errors and noise-induced barren plateaus~\cite{wang,fontana2021evaluating,brandhofer2021error}. VQE has been experimentally demonstrated for a variety of problem instances~\cite{vqe,o2016scalable,shen2017quantum,kandala,santagati2018witnessing,google2020hartree}, but has so far been inadequate in producing results for any problem that could be considered classically intractable~\cite{brandhofer2021error,lee2022there}. Solving larger problems on noisy devices using VQE is difficult because variational quantum algorithms have a tension to overcome: the necessary precision in practical application requires encoded trial states with large ground state overlap for the problem Hamiltonian, but better ans\"atze have high circuit depth and are thus typically incompatible with NISQ devices due to the high amount of noise introduced in the energy expectation value estimate $\langle H \rangle$.

Here we focus on the recently introduced quantum computed moments (QCM) approach, which seeks to mitigate this trade-off by improving the ground state energy estimate via the quantum computation of higher order moments of the Hamiltonian~\cite{hv2020} and application of powerful results from Lanczos expansion theory~\cite{pexp,general,fz}. By leveraging information provided by the moments of the Hamiltonian rather than relying on a fully converged trial state, the QCM method has already been shown to improve on the variational estimate and compensate for a low-depth ansatz choice in several applications~\cite{hv2020,mj-chem}.

This paper explores and develops the QCM method as a noise-robust heuristic. The relatively low level of sensitivity to noise was observed as an unexpected feature of the QCM technique when applied to low depth circuits~\cite{hv2020}. Further work~\cite{mj-chem} achieved high precision for small-scale molecular problems using QCM on a noisy device, in conjunction with error mitigation strategies, in a study on the ability of the moments approach to correct for the missing description of electronic correlations in the VQE trial state ansatz. However, the primary focus of this paper is on investigating the robustness of the QCM method to deep trial state circuits and the subsequent outcome of a novel noise-robust heuristic. Here, in the context of quantum magnetism, we investigate how QCM inherently compensates for noise introduced in deep, expressive circuits. We show that, remarkably, the QCM energy estimate consistently achieves good accuracy even when conventional approaches (based on $\langle H \rangle$ alone) break down. Our
primary result demonstrates this robustness for a 20 qubit example of the Heisenberg spin model encoded on an IBM Quantum device with trial state circuits of increasing depth up to hundreds of CNOTs. The QCM ground state energy estimate maintains $\lesssim$ 10$\%$ approximation error well into the deep trial state circuit regime where errors overwhelm both the typical VQE estimate and other comparable moments-based formulae. Based on the moments-derived `high-temperature' limit of the model corresponding to the maximally mixed trial state, these observations point towards a novel robust heuristic for such problems. 

\begin{figure*}[ht!]
\includegraphics[width=0.9\linewidth]{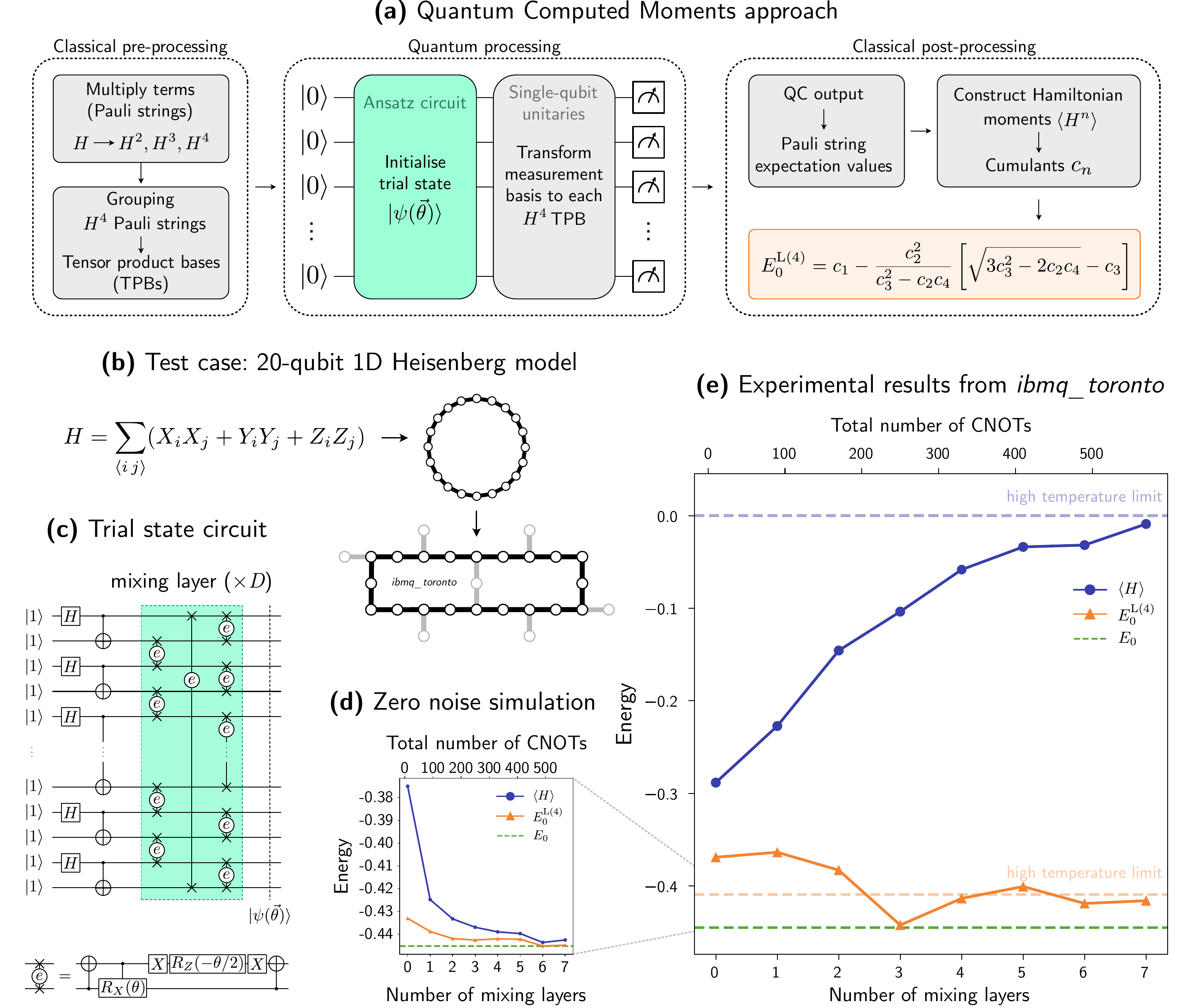}
\centering
\caption{
Application of the QCM approach to solving a 20-qubit quantum Heisenberg model. \textbf{(a)}~Overview of the QCM approach. A trial state $\ket{\psi(\vec{\theta})}$ with optimised parameters $\vec{\theta}$ is measured in the relevant tensor product bases to produce Hamiltonian moments $\left\langle H^n \right\rangle$ 
which are then combined to form the Hamiltonian ground state energy estimate $E_0^{\rm{L}(4)}$ at an $n=4$ cumulant truncation.
\textbf{(b)}~20-qubit 1D Heisenberg model Hamiltonian with uniform couplings and the corresponding mapping onto the 27-qubit \textit{ibmq\_toronto} superconducting device. \textbf{(c)}~We use the 1D RVB trial state circuit from~\cite{seki2020} constructed from a singlet state and $D$ mixing layers of parameterised eSWAP gates. \textbf{(d)}~Results from zero noise simulations of RVB circuits with up to $D=7$ mixing layers and near-optimal parameters $\vec{\theta}$, comparing ground state energy estimates of Hamiltonian expectation value $\left\langle H \right\rangle$ and $E_0^{\rm{L}(4)}$. \textbf{(e)}~Ground state energy estimate comparison of results from \textit{ibmq\_toronto} after running the same near-optimal circuits (8 trial states $\times$ 1792 TPB measurements $\times$ 8192 shots). Dotted lines correspond to ``high-temperature'' limits, i.e. the values of $\langle H \rangle$ (blue) and $E_0^{\rm{L}(4)}$ (orange) evaluated with respect to the maximally mixed state.
}
\label{Fig1}
\end{figure*}

This work is organised as follows. First, in Section~\ref{sec2}, we give a brief overview of the QCM method for estimating the ground state energy of a Hamiltonian system. Next, we apply the method to the 20-qubit Heisenberg model on IBM Quantum computer hardware, demonstrating the remarkable resilience of the technique to quantum error. This is our main result, summarised in Figure~\ref{Fig1}. We then provide a theoretical analysis of this apparent noise robustness via an analytic model, showing the error cancellation explicitly in a simplified model with Heisenberg-like structure under global white noise. In Section~\ref{sec3}, we verify the versatility of the QCM approach, studying its robustness to noise for an ensemble of random instances of quantum magnetism Hamiltonians on real devices and under more realistic noisy simulations. We build on the idea of using QCM as a quantum heuristic for ground state energy problems by benchmarking it against a `high-temperature' limit from classically computed moments of the maximally mixed state. Finally, in Section~\ref{sec4}, we conclude with a discussion and summary of our results.

\section{Noise robustness in the QCM approach} \label{sec2}
The variational approach to Hamiltonian problems is a familiar starting point in quantum computing. Given a trial state $\ket{\psi(\vec{\theta})}$, parameterised by $\vec{\theta}$, the expectation value of the Hamiltonian $\langle H\rangle \equiv \bra{\psi(\vec{\theta})} H \ket{\psi(\vec{\theta})}$ is an upper bound to the true ground state energy $E_{\rm 0}$ and can be minimised with respect to $\vec{\theta}$. On a quantum computer, the trial state is encoded by a quantum circuit parameterised by $\vec{\theta}$ and the output sampled to produce estimates of $\langle H\rangle$ (the first Hamiltonian moment), which are then fed into a classical optimisation loop to minimise over $\vec{\theta}$~\cite{vqe,qaoa}. In this optimisation loop, it is the cost function evaluation step -- the preparation and measurement of $\langle H \rangle$ -- that is the primary action of the quantum computer, where variational quantum algorithms are thought to have an edge over their classical counterparts. It was realised some time ago that quantum computers may also be useful in the determination of Hamiltonian moments~$\langle H^n \rangle$~\cite{duan, jones}, and recently there have been a number of developments on this theme, both theoretically and experimentally~\cite{hv2020,mj-chem,kowalski2020quantum,seki2021quantum,suchsland2021algorithmic,momentsreview}.

Our approach provides a moments-based correction to the ground state energy estimate, and is based on the QCM framework~\cite{hv2020} in the context of Lanczos expansion theory for Hamiltonian systems~\cite{pexp}. In particular, we focus on the analytical expression for the order $\langle H^4\rangle$ estimate of the ground state energy, $E_0^{\rm L(4)}$, which was derived from Lanczos expansion theory in~\cite{general}: 
\begin{equation}
E_{\rm 0} \sim E_{\rm 0}^{\rm{L}(4)} = c_1 - \frac{c_2^2}{c_3^2 - c_2 c_4} \left[\sqrt{3 c_3^2 - 2 c_2 c_4} - c_3\right], \label{formula}
\end{equation}
where $c_n$ are the cumulants associated with the Hamiltonian moments $\langle H^n\rangle$:
\begin{equation}
c_n = \langle H^n\rangle - \sum_{p=0}^{n-2}\binom{n-1}{p}c_{p+1} \langle H^{n-1-p}\rangle .
\end{equation}
%
Eq~\eqref{formula} is an exact diagonalisation via an infimum theorem~\cite{fz} to working order in the moments, and is applicable to extensive and non-extensive systems~\cite{nonext}. For a given $\ket{\psi(\vec{\theta})}$ with some ground state overlap, $E_0^{\rm L(4)}$ takes the form of a first order variational estimate ground state energy, $c_1=\langle H \rangle$, plus a correction which employs higher order correlations of the system encapsulated in the $c_n$. In the context of quantum computation, this is a critical feature of the QCM technique, given that the preparation of a trial state and computation of the first order ground state energy estimate $\langle H \rangle$ is the key quantum step in hybrid variational quantum algorithms such as VQE.

An outline of the QCM approach is summarised in Figure~\ref{Fig1}(a). First, a trial state is prepared on the quantum computer from a well-chosen ansatz circuit. When the Hamiltonian is expressed in spin operator form, the expectation value of $H$ can be determined term-by-term, where each term is a weighted “Pauli string” whose expectation value can be determined by applying corresponding change of basis unitaries to each qubit followed by measurement. Pauli strings that qubit-wise commute with one another can be grouped into the same tensor product basis (TPB), as to measure their expectation values simultaneously. Thus to measure the higher order moments, one can multiply out the terms in the Hamiltonian to obtain expressions for $H^2$, $H^3$ and $H^4$, and apply the same procedure to obtain the expectation values of the resulting Pauli strings. It turns out that, though the amount of terms in these higher order expressions scales polynomially with the number of terms in $H$, the actual number of runs required on the quantum computer (one for each TPB), scales only logarithmically with the terms in $H$~\cite{hv2020}. Once the moments up to $\langle H^4 \rangle$ are computed, they are combined to form $E_0^{\rm L(4)}$, the improved ground state energy estimate.

The improvement of QCM over the variational result was explicitly demonstrated in~\cite{hv2020} for quantum magnetism systems of up to 25 qubits with respect to the simple N\'eel state, and for chemistry problems in~\cite{mj-chem} producing results beyond the Hartree-Fock variational limit. These previous results show that QCM is a robust method for determining better estimates to the ground state energy of a system when using relatively simple trial states, e.g. when ground state overlap is sacrificed in favour of having shallower circuits, or in problems where the ansatz circuit does not fully take into account the dynamics of the system at hand.

QCM produces the quantity $E_{\rm 0}^{\rm{L}(4)}$ from additional measurements on the quantum computer (polynomial in the problem size), and this quantity has a less stringent requirement of trial state complexity than the variational estimate, $\langle H \rangle$. In this work, however, we instead focus on another property of $E_{\rm 0}^{\rm{L}(4)} $ observed in the applications to date~\cite{hv2020,mj-chem} -- that the expression possesses robustness to noise. We will demonstrate that $E_0^{\rm L(4)}$, as a particular combination of moments, is effective in producing energy estimates even when errors accumulated in deep trial state circuits overwhelm the variational calculation. We start with a demonstration of the QCM procedure applied to very deep quantum circuits on a real device, from which $E_{\rm 0}^{\rm{L}(4)}$ surprisingly salvages a meaningful result. We will then investigate this noise robustness property with a simple theoretical model.

\subsection{Demonstration of QCM for a 20-qubit Heisenberg model}\label{sec:20q}
A summary and demonstration of the QCM method is shown in Figure~\ref{Fig1}. For our context, we take the Heisenberg spin model with Hamiltonian given by: 
\begin{equation}
H = \frac{1}{4q} \sum_{\langle i\,j\rangle} \left(J_{ij}^{(x)} X_i X_j + J_{ij}^{(y)} Y_i Y_j + J_{ij}^{(z)} Z_i Z_j\right),
\end{equation}
where the sum is over a problem graph defined by the vertices (qubits) $ {i =1...q}$, edges connecting qubits $\{\langle i\,j\rangle\}$, and couplings $J_{ij}^{(s)}$ along each edge ($s = x, y, z$). Here we consider nearest-neighbour linear-lattices with periodic boundary conditions. The uniform coupling case $J_{ij}^{(x)} = J_{ij}^{(y)} = J_{ij}^{(z)}$ is the well known Heisenberg model, for which the exact ground state has been studied numerically for decades and is often used for testing new approaches. In contrast to the N\'eel trial state of~\cite{hv2020}, here we explicitly focus on the deep quantum circuit regime. As our class of trial states we use the resonating-valence-bond (RVB) circuit from Seki \textit{et al.}~\cite{seki2020} shown in Figure~\ref{Fig1}(c), where the trial state over $q$ qubit couplings is built from $D$ mixing layers of $q$ exponential SWAP (eSWAP) gates, each parameterised by an angle $\theta_i$ with a total set of $D\times q$ parameters $\vec{\theta} = (\theta_1,\theta_2,...,\theta_{Dq-1},\theta_{Dq})$.  The procedure is defined as follows: for a trial state defined over $D$ mixing layers, $|\psi(\vec{\theta})\rangle$, we compute $\langle H\rangle_{\vec{\theta}} \equiv \langle \psi(\vec{\theta}) |H|\psi(\vec{\theta})\rangle$ under zero noise simulation and minimise over the parameter set $\vec{\theta}$. At the minimum point we obtain the variational estimate $\langle H\rangle_{\vec{\theta}^*}$ where $\vec{\theta}^*$ is a corresponding near-optimal parameter set. For that trial state configuration we then compute the moments $\langle H^n \rangle_{\vec{\theta}^*}$ up to $n=4$ which are fed into the formula for the fourth order Lanczos ground state energy estimate, $E_{\rm 0}^{\rm{L}(4)}$. The zero noise simulation for the uniform coupling case, shown in Figure~\ref{Fig1}(d), shows that the moments-based estimate $E_0^{\rm{L}(4)}$ provides a systematic correction to the minimum $\langle H\rangle_{\vec{\theta}^*}$, compensating for the degree of overlap with the ground state. The circuits corresponding to the parameter sets $\vec{\theta}^*$ were then compiled and run on the IBM Quantum device {\em ibmq\_toronto}, with results shown in Figure~\ref{Fig1}(e). Hamiltonian moments up to $n=4$ were computed from TPB measurements as per the heuristic method outlined in~\cite{hv2020}. The $282\:796$ terms in $H^4$ were reduced to $1792$ TPB terms that are simultaneously measurable on the quantum computer. Moments and associated cumulants were then constructed from these measurements ($1792\times 8192$ shots).

Figure~\ref{Fig1}(e) shows the stark difference between the results obtained from $E_{\rm 0}^{\rm{L}(4)}$ and $\langle H \rangle_{\vec{\theta}^*}$ when running these circuits on real noisy hardware, as opposed to zero noise simulation in Figure~\ref{Fig1}(d). Due to the errors in the device, the variational results $\langle H \rangle_{\vec{\theta}^*}$ start at a higher energy and move rapidly away from the exact ground state as the trial state complexity builds, converging towards the high-temperature limit of $\langle H \rangle$ corresponding to the maximally mixed state. As we will see in Section~\ref{ensemble}, this behaviour is exactly as expected for $\langle H \rangle$, and indeed the energy estimate error on $\langle H \rangle$ follows a simple ``NISQ \textit{prima facie}'' scaling with the number of CNOTs at a $\sim1\%$ error rate. In contrast, the results from $E_{\rm 0}^{\rm{L}(4)}$, although pressured by a lower high-temperature limit, remain in proximity to the true ground state even for high depth trial states with hundreds of CNOTs applied. We note that all results shown are the raw data obtained from the device -- no attempt at error mitigation has been made as yet. We note that $E_{\rm 0}^{\rm{L}(4)}$ is an exact diagonalisation of the Hamiltonian to working moment order, a fact that may be responsible for the noise-robustness property. To shine some light on this, we investigate this error filtering effect for a simple model.

\subsection{Analysis of Heisenberg-like model under global white noise}\label{sec2:toy}
Here we seek to understand the apparent robustness of $E_0^{\rm L(4)}$ versus the variational ground state estimate $\langle H \rangle$ when inputs are derived from noisy quantum computations. We will also compare our Lanczos expansion theory based method to a similar method derived from the $t$-expansion~\cite{horn1984t}, from which the ground state energy of a quantum many-body system can be extrapolated in terms of Hamiltonian moments in a number of ways~\cite{stubbins1988methods} -- the most well-known of these being the connected moments expansion (CMX)~\cite{cmx}:
\begin{equation}
E_0^{\rm CMX} = c_1 - \frac{c_2^2}{c_3}-\frac{1}{c_3}\frac{\left(c_2c_4-c_3^2\right)^2}{c_3c_5-c_4^2}-\ldots.
\end{equation}
The CMX is a suitable benchmark for our Lanczos derived moment method, as it has seen recent interest in a quantum computing context~\cite{kowalski2020quantum,seki2021quantum}. When truncated to $K$ terms, the CMX involves Hamiltonian moments up to order $(2K-1)$. For a generous comparison with $E_0^{\rm L(4)}$ (at 4th order in the moments), we take the three term truncation of the CMX ground state energy (to 5th order in the moments), denoted as $E_0^{\rm CMX(5)}$ to remain consistent with our notation.

We will now construct an analytic model to get a flavour of how $E_0^{\rm L(4)}$, and hence the QCM approach, deals with noise. In our model, we consider global white noise, an assumption which a recent result~\cite{dalzell2021random} suggests is not entirely unrealistic when considering how errors propagate through deep, highly entangling quantum circuits. Under white noise, Hamiltonian moments transform as:
\begin{equation}\label{noisymoment}
    \langle H^k \rangle \rightarrow \langle H^k \rangle_{\rm noise} = (1-p)\langle H^k \rangle + \frac{p}{N}{\rm tr}(H^k),
\end{equation}
where $p\in[0,1]$ is the noise parameter and $N=2^q$ is the system Hilbert space size. 
In general, plugging these noisy moments into the formulae for $E_0^{\rm L(4)}$ and $E_0^{\rm CMX(5)}$ results in ungainly expressions with no obvious structure with respect to the noise parameter. However, we can make some simplifications to the Hamiltonian in order to study the noise robustness. We consider a three-level toy model that mimics the low energy structure of the 1D Heisenberg model. We take a Hamiltonian with ground state energy $E_0$ and write the excited gap structure as:
\begin{align}
E_1 = E_0 + \Delta_1, && E_2 &= E_0 + \Delta_1 + \Delta_2,
\end{align}
with $\Delta_1>0$ and $\Delta_2>0$. We define the following energy gap ratio:
\begin{equation}
    R = \frac{E_2-E_1}{E_1-E_0}=\frac{\Delta_2}{\Delta_1}.
\end{equation}
For small values of $R$, the system resembles the low-lying states of the $2L$-site 1D Heisenberg model as $L$ increases.
The convergence properties of the original Lanczos algorithm are governed by the gap of the systems~\cite{witte}, namely that more rapid convergence occurs when the $E_1-E_0$ gap is larger (relative to the entire spectral width). Studying the behaviour of this three-level system for small values of $R$ ($\Delta_1 \gg \Delta_2$) therefore also leaves us in the regime of rapid Lanczos convergence, where the potential noise robustness can be observed in full effect.

To study the effect of noise in the deep trial state limit, we take moments evaluated with respect to the exact ground state, i.e.
\begin{equation}
    \langle H^k \rangle = E_0^k.
\end{equation}
Noting that the first excited state of the $2L$-site 1D Heisenberg model is threefold degenerate, the expressions for the moments in the presence of noise from Eq~\eqref{noisymoment} now become:
\begin{multline}
    \langle H^k \rangle_{\rm noise} = (1-p)E_0^k \\
    +\frac{p}{5}\left(E_0^k + 3(E_0 + \Delta_1)^k + (E_0 + \Delta_1 + \Delta_2)^k\right).
\end{multline}
Substituting these noisy moments and expanding in $R$ yields the following transformations of each estimate (variational, CMX and Lanczos) with respect to the exact ground state under global white noise:
\begin{flalign}
    \langle H \rangle &\rightarrow\; E_0 + \frac{p}{5}(4\Delta_1 + \Delta_2), &\\
    E_0^{\rm{CMX}(5)} &\rightarrow\; E_0 + \frac{16 p^3 (4\Delta_1+\Delta_2)}{125 - 300 p + 240 p^2}\notag&\\ 
     &+ R\left[\Delta_2\frac{375 p - 3900 p^2 + 2880 p^3 + 2304 p^4}{16 (25 - 60 p + 48 p^2)^2}\right]\notag &\\
    &+ \mathcal{O}\left(R^2\right),  &\\
    E_0^{\rm{L}(4)} &\rightarrow\; E_0 + R\left[\Delta_2\frac{15 p - 24 p^2}{8 (5 - 4 p)^2}\right]+\mathcal{O}\left(R^2\right).&
\end{flalign}
The key observation is that at zeroth order in $R$, the dependence on the noise parameter $p$ has cancelled out in $E_0^{\rm L(4)}$, unlike in $\langle H \rangle$ and $E_0^{\rm{CMX}(5)}$ which have a linear effect in $p$. Additionally, the coefficient of the $R$ term in $E_0^{\rm{L}(4)}$ remains relatively constant with $p$ and close to zero up until $p\approx0.7$, after which it deviates from zero by no more than $9\%$ of $\Delta_2$. Hence, in this limit we see how the expression $E_0^{\rm L(4)}$ possesses an inherent robustness in comparison to the VQE and CMX estimates.

It has been shown previously in~\cite{nonext} that the Lanczos expansion to fourth order for the harmonic $N$-boson model with respect to the Hartree trial state diagonalises the system exactly, so this result showing the exact cancellation of $E_0^{\rm{L}(4)}$ down to $E_0$ is perhaps not surprising in the small $p$ regime. In any case, we conclude for this analytic model that the $E_0^{\rm L(4)}$ expression is remarkably robust to $p$, suggesting that with a well-chosen trial state ansatz we can recover the result $E_0^{\rm{L}(4)}\rightarrow E_0$ under a considerably high degree of noise on the individual moments. This result validates our observation for a large problem instance on a real quantum computer in Figure~\ref{Fig1}, where we see an apparent resilience to errors for deep ansatz circuits with the computed ground state energy estimate approaching the exact solution despite an abundance of noise. We are now motivated to study this approach for an ensemble of random instances of the Heisenberg Hamiltonian and noise models, and ultimately test on physical QC devices.
\begin{figure*}[ht!]
\includegraphics[width=\linewidth]{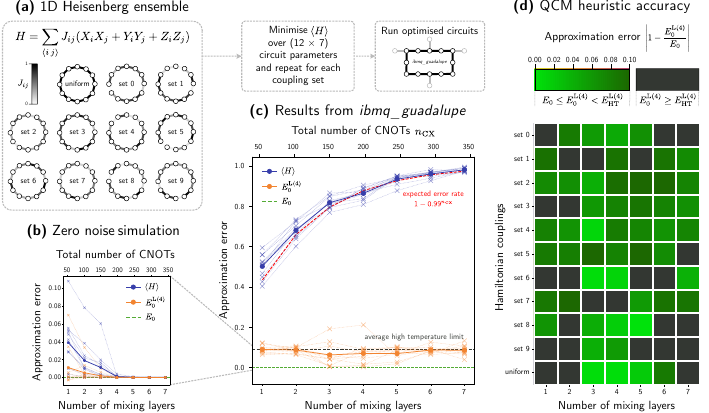}
\centering
\caption{Results from an ensemble of 12-qubit 1D Heisenberg Hamiltonians with random couplings $J_{ij} \in [0,1]$ simulated both with zero noise and on a real quantum processor. The trial states used are 12-qubit 1D RVB circuits up to $D=7$ layers optimised under zero noise with respect to the various Hamiltonians. \textbf{(a)}~Diagrams of the uniform coupling case and the 10 random coupling instances of the Hamiltonian, and mapping onto the 16-qubit \textit{ibmq\_guadalupe} device. \textbf{(b)}~Approximation error $|1-\frac{E}{E_0}|$ comparison between energy estimates $\left\langle H \right\rangle$ and $E_0^{\rm{L}(4)}$ under zero noise simulation and \textbf{(c)}~on the \textit{ibmq\_guadalupe} quantum processor. Dotted lines represent the individual Hamiltonians, and solid lines show the average over the ensemble. The average \textit{ibmq\_guadalupe} CNOT fidelity is used to calculate the NISQ \textit{prima facie} expected total error of each circuit, shown as a dashed red curve. \textbf{(d)}~Heat map of $E_0^{\rm L(4)}$ approximation error for each Hamiltonian in the ensemble. Each square corresponds to a single run (990 basis measurements $\times$ 8192 shots) on the quantum processor. Each green square represents an obtained value of $E_0^{\rm L(4)}$ that falls below its respective high-temperature limit, $E_{\rm HT}^{\rm L(4)}$, which can be computed efficiently.}
\label{Fig2}
\end{figure*}

\section{Versatility of the QCM approach} \label{sec3}
In this section, we study the effectiveness of the QCM approach as a heuristic for solving ground state energy problems, both in its accuracy and its robustness to noise on real quantum hardware. By considering a broader range of Hamiltonians and more realistic error models, we validate the results in the previous section and showcase the general applicability of the method.

 \subsection{Application of QCM to an ensemble of 1D Heisenberg models} \label{ensemble}
Here we apply the QCM method to different Hamiltonians and discuss the prospect of the approach as a noise robust quantum heuristic. Figure~\ref{Fig2} shows the experiment from Section~\ref{sec:20q} repeated several times on the \textit{ibmq\_guadalupe} device, each for a random coupling instance of a 12-qubit 1D Heisenberg Hamiltonian. The ensemble of cases studied and their mappings onto the device are shown in Figure~\ref{Fig2}(a). Going up to $D=7$ mixing layers, near-optimal trial state circuits of increasing depth were found for each Hamiltonian in the ensemble by minimising $\langle H \rangle$ with respect to parameters in the RVB ansatz~\cite{seki2020} under zero noise simulation. The ground state energy approximation error $\left| 1-E/E_0 \right|$ of the variational estimate $\langle H \rangle$ and the QCM estimate $E_0^{\rm L(4)}$ for each circuit under this zero noise simulation is shown in Figure~\ref{Fig2}(b). Here, the trial state energies $\langle H \rangle$ are on average within $5\%$ of the ground state energy, and those with $D\geq4$ mixing layers rapidly converging to the exact ground state. Figure~\ref{Fig2}(c) shows the approximation error of each energy estimate after running the circuits on the noisy \textit{ibmq\_guadalupe} device (990 TPB measurements $\times$ 8192 shots $\times$ 7 mixing layers $\times$ 11 Hamiltonians). Also shown is the NISQ \textit{prima facie} expected error rate of each circuit, $1-(1-\varepsilon_{\rm CX})^{n_{\rm CX}}$, calculated from the total number of CNOTs $n_{\rm CX}$ and the average CNOT error $\varepsilon_{\rm CX}$ on the device at the time of the experiments (typically~$\sim 1\%$).

Across the ensemble, we see broadly the same result as in Figure~\ref{Fig1} -- that the QCM approach consistently offers a remarkably noise-robust correction (up to $\sim 90\%$) to the variational ground state energy estimate across all random Heisenberg models studied, even when using trial state circuits with hundreds of CNOT gates. We once again note that the results shown are raw data from the device, where no error mitigation techniques have been applied (but we expect error mitigation to improve the results).
We see that the approximation error of $\langle H \rangle$ closely tracks the expected error rate scaling with the number of CNOTs in each variational circuit, implying that with increasing circuit complexity, typical error rates on NISQ devices will always be a barrier to meaningful application of VQE. However, we see in the behaviour of $E_0^{\rm L(4)}$ that the NISQ \textit{prima facie} barrier may be broken when taking into account higher order moments in the quantum computed energy estimate.
These results provide strong evidence that in the calculation of moments on quantum computer, errors entering into the trial state preparation can be relatively high and yet the  moments extracted produce high quality estimates from $E_{\rm 0}^{\rm{L}(4)}$. Putting aside shot noise, errors in the quantum computation of the moments only ever accumulate in the trial state itself. One might expect the broad traction that the Lanczos procedure obtains from a given trial state smooths the sensitivity to error fluctuations, but the analytic model analysis also indicates cancellations occur in the formula for $E_{\rm 0}^{\rm{L}(4)}$ (that cannot occur for $\langle H \rangle$ alone) that possibly dominate the robustness behaviour.

We note that, in Figure~\ref{Fig1}(e) and Figure~\ref{Fig2}(c), as circuit depth increases on the real device, each energy estimate approaches their corresponding high-temperature limit, i.e. the energy estimate evaluated with respect to the maximally mixed state. We shall denote the high-temperature limit of the QCM estimate $E_{0}^{\rm L(4)}$ as $E_{\rm HT}^{\rm L(4)}$. In all of the noisy cases studied, $E_{\rm HT}^{\rm L(4)}$ seemingly outperformed the noisy variational estimate $\langle H \rangle$. This is particularly interesting because $E_{\rm HT}^{\rm L(4)}$ can be computed solely from the identity term coefficients of $\{H, H^2, H^3, H^4\}$, i.e. the high-temperature limit does not require any quantum computation and can be obtained efficiently classically. This leaves us with a classical heuristic that is more accurate than the variational estimate on current NISQ devices with no error mitigation. We can use this maximally mixed $E_{\rm HT}^{\rm L(4)}$ result to benchmark the QCM results, to define a quantum version of this heuristic.

\begin{figure*}[ht!]
\includegraphics[width=0.9\linewidth]{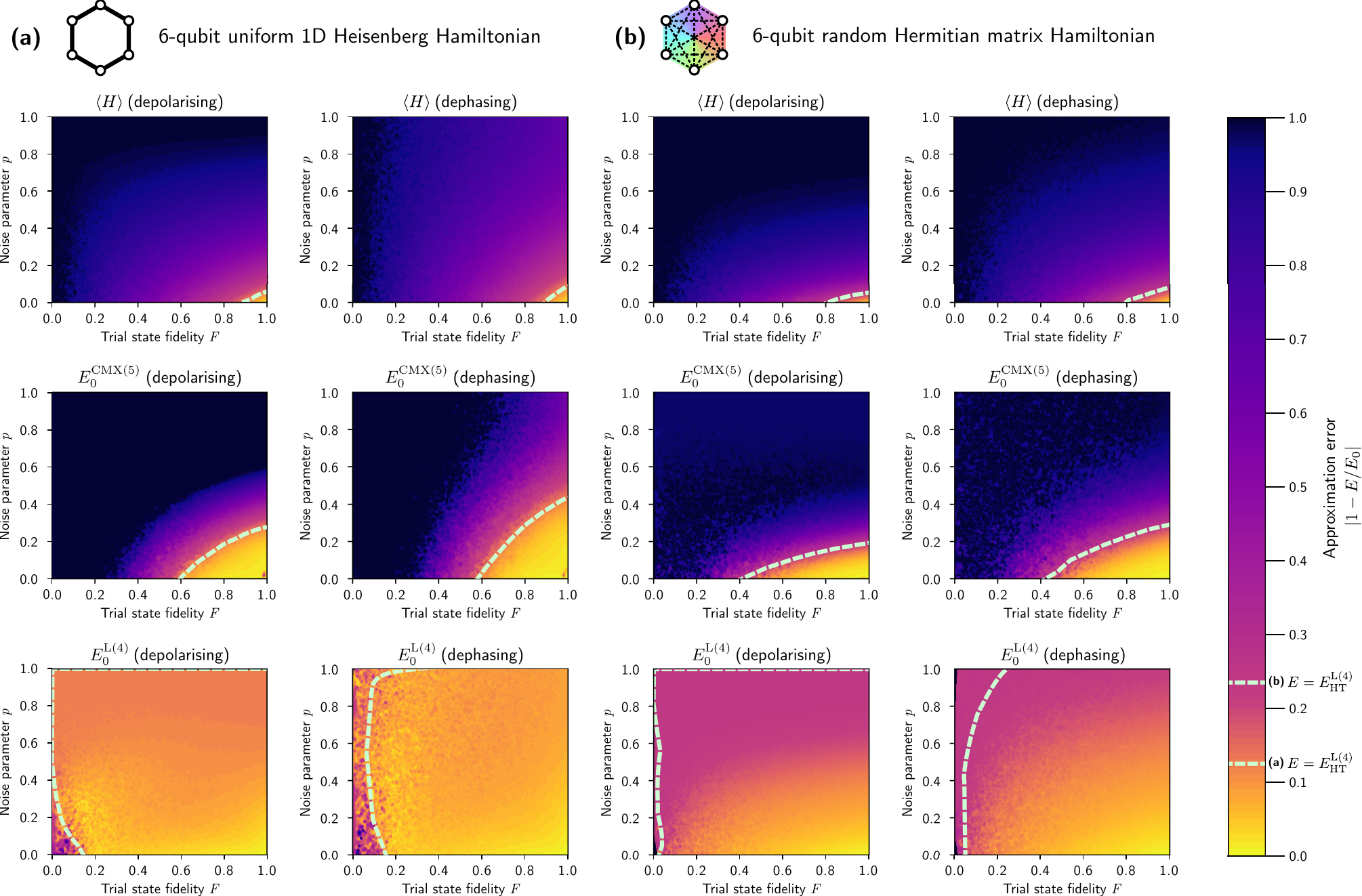}
\centering
\caption{Noise robustness analysis of the energy estimates $\left\langle H \right\rangle$, $E_0^{\rm{CMX}(5)}$ and $E_0^{\rm{L}(4)}$ for \textbf{(a)}~a uniform 1D Heisenberg Hamiltonian and \textbf{(b)}~a random Hermitian matrix Hamiltonian under simple depolarising (${\rho} \mapsto (1-\frac{3p}{4}){\rho} + \frac{p}{4}(X{\rho}X+Y{\rho}Y+Z{\rho}Z)$) and dephasing (${\rho} \mapsto (1-\frac{p}{2}){\rho} + \frac{p}{2}Z{\rho}Z$) noise models. Various trial states are generated, and the approximation error $|1-\frac{E}{E_0}|$ of the three energy estimates are plotted against noise parameter $p$ and zero noise state fidelity $F = \braket{\phi_0|{\rho}|\phi_0}$ with respect to $\ket{\phi_0}$, the ground state of the Hamiltonian. The trial states ${\rho}=\ket{\phi_\rho}\!\bra{\phi_{\rho}}$ are chosen by applying random rotations from the ground state as $\ket{\phi_\rho} = e^{-i\varepsilon M}\ket{\phi_0}$, where $M$ is a random Hermitian matrix sampled from the Gaussian unitary ensemble and $\varepsilon$ is a parameter varied from zero until the $F$-axis is filled out. Dashed pale green lines show the boundaries on each plot where the approximation error is equal to that of $E_{\rm HT}^{\rm L(4)}$, the high-temperature limit of $E_0^{\rm L(4)}$.}
\label{Fig3}
\end{figure*}

Figure~\ref{Fig2}(d) shows where the measured $E_{0}^{\rm L(4)}$ on the NISQ device outperformed the $E_{\rm HT}^{\rm L(4)}$ estimate of the ground state energy, allowing useful information to be obtained from the quantum computer. We see that for all test cases studied, there exists a regime on present-day NISQ devices in which the QCM heuristic has an advantage over the classical benchmark. The efficiently computable high-temperature limit $E_{\rm HT}^{\rm L(4)}$ could thus serve to check the validity of the ground state energy estimate when performing the QCM calculation, as it falls out of the result for $E_{0}^{\rm L(4)}$ with no extra computation required.


\subsection{General behaviour of QCM}

The experimental results for the various quantum magnetism models shown in Figure~\ref{Fig2} and the analytic model analysis from Section~\ref{sec2:toy} indicate that the noise robustness of QCM should persist for an arbitrary choice of Hamiltonian. We will now investigate this general behaviour by looking at QCM under more realistic error models with respect to both the Heisenberg model and more general random Hamiltonian instances.

In Figure~\ref{Fig3}, we investigate the versatility of the QCM approach by considering a random $2^6\times2^6$ Hermitian matrix as a Hamiltonian, in addition to the uniform Heisenberg model from earlier on 6 qubits. Energy estimates $\langle H \rangle$, $E_0^{\rm{L}(4)}$ and $E_0^{\rm{CMX}(5)}$ are computed with respect to randomly generated trial states~${\rho}$ under noisy simulations of depolarising and dephasing error models, which transform the states via a noise parameter $p \in [0,1]$ as:
\begin{align*}
&\text{depolarise:\hspace{-0.3cm}} &{\rho} \mapsto &(1-\frac{3p}{4}){\rho}+\frac{p}{4}(X{\rho}X+Y{\rho}Y+Z{\rho}Z),&\\
&\text{dephase:} &{\rho} \mapsto &(1-\frac{p}{2}){\rho} + \frac{p}{2}Z{\rho}Z.&
\end{align*}
These models give a reasonable approximation to the type of error seen in deep circuits on a real quantum computer, and on a present-day NISQ device with $\sim1\%$ CNOT error, one might expect $p\approx0.3-0.4$ for a typical VQE circuit.

\begin{figure*}[ht!]
\includegraphics[width=0.9\linewidth]{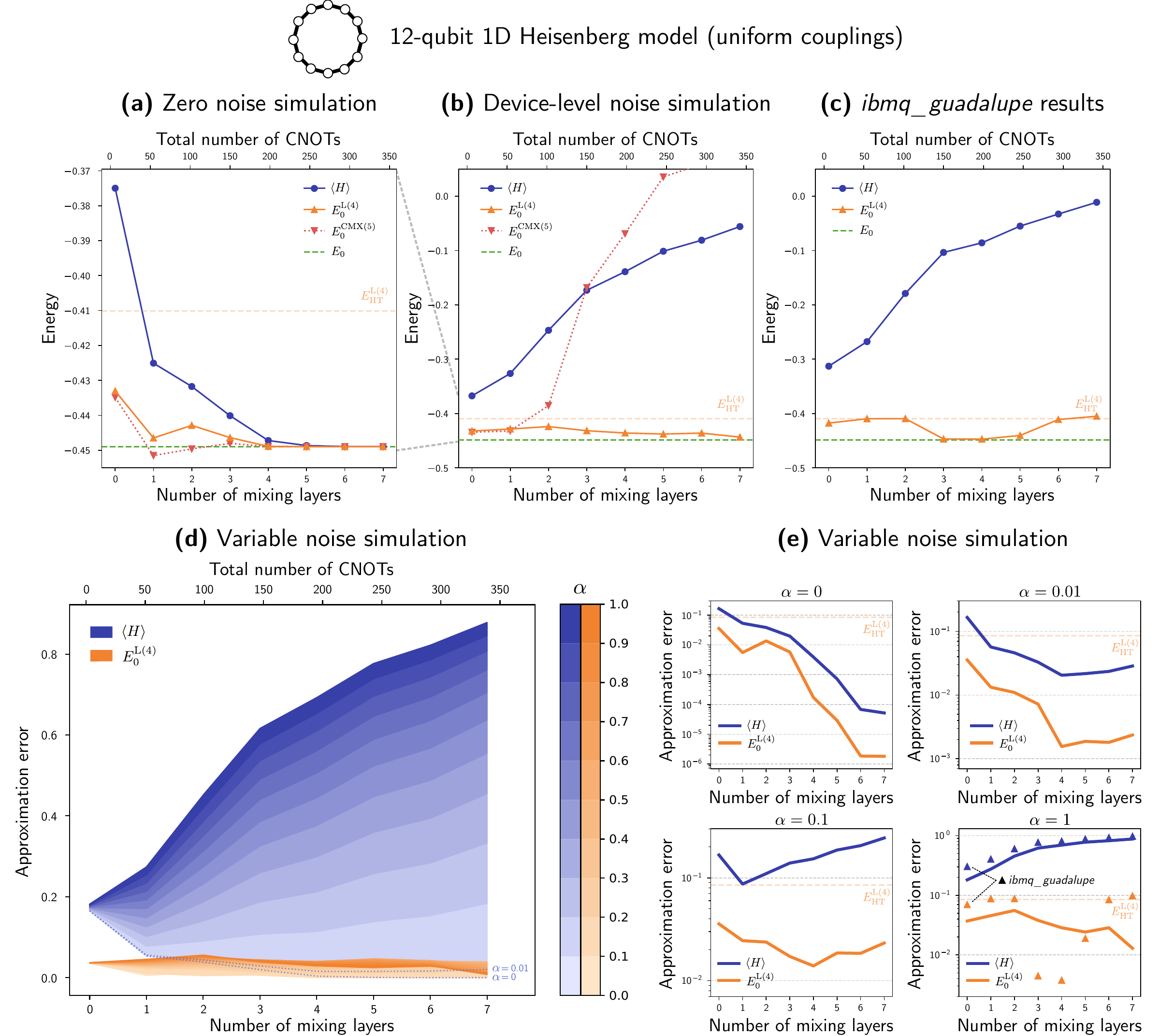}
\centering
\caption{Further robustness analysis of the energy estimates $\langle H \rangle$, $E_0^{\rm{CMX}(5)}$ and $E_0^{\rm{L}(4)}$ using a circuit-based error model and optimised RVB trial state circuits up to $D=7$ layers for a 12-qubit 1D Heisenberg Hamiltonian. \textbf{(a)}~Zero noise simulation of the energy estimates, \textbf{(b)}~device-level noisy simulation incorporating readout, thermal relaxation and depolarising error to match calibration data obtained from the \textit{ibmq\_guadalupe} backend, and \textbf{(c)}~comparison with experimental results from the device. All readout and gate errors in the noisy simulation are multiplied by a factor $\alpha$, and \textbf{(d)}~shows the approximation error of $\langle H \rangle$ and $E_0^{\rm{L}(4)}$ under the noisy simulation, varied from device-level ($\alpha=1$) down to zero noise ($\alpha=0$). For clarity, \textbf{(e)}~shows instances of these same simulations on a log scale for a range of $\alpha$ values, overlaid with results from \textit{ibmq\_guadalupe} at $\alpha=1$.}
\label{Fig4}
\end{figure*}

The horizontal axis, $F=\braket{\phi_0|{\rho}|\phi_0}$, of each plot in Figure~\ref{Fig3} is a measure of the closeness of each trial state to the true ground state~$\ket{\phi_0}$ at zero noise. The vertical axis, $p$, represents the extent to which each trial state has been scrambled by noise. The typical trajectory of a variational quantum algorithm would thus be from left to right as trial state parameters change, and from bottom to top as more circuit layers are added. Along this entire trajectory, we see that $E_0^{\rm{L}(4)}$ much more closely approximates the ground state energy of each system than $\langle H \rangle$ and $E_0^{\rm{CMX}(5)}$. The noise robustness of the QCM approach persists in general, whereas this behaviour is not observed in the other two energy estimates. Additionally, in the low noise regime, $E_0^{\rm{L}(4)}$ still gives a more accurate estimate of the true ground state energy than $\langle H \rangle$ and $E_0^{\rm{CMX}(5)}$ at lower values of $F$, supporting the previous results from~\cite{hv2020} that indicate the QCM method's effectiveness when less complicated ansatz circuits are used.

Also shown in Figure~\ref{Fig3} are dashed lines corresponding to the points at which each of $\langle H \rangle$, $E_0^{\rm CMX(5)}$ and $E_0^{\rm L(4)}$ are equal to $E_{\rm HT}^{\rm L(4)}$, the high-temperature limit of $E_0^{\rm L(4)}$, which is classically efficient to compute. These outline the regions inside which there is some advantage in estimating the ground state energy using a quantum computer rather than the na\"ive classical calculation of $E_0^{\rm L(4)}$ with maximally mixed moments. For the Hamiltonians and error models studied, these regions take up most of the $E_0^{\rm L(4)}$ plots, and are akin to the green squares displayed in Figure~\ref{Fig2}(d), where a similar analysis was performed on real device data.

The same picture as in Figure~\ref{Fig3}(b) was seen when averaging over thousands of other randomly generated Hamiltonian instances in simulations of up to 10 qubits ($2^{10}\times2^{10}$ random Hermitian matrices). The striking error robustness of $E_0^{\rm L(4)}$ under these random Hermitian matrix Hamiltonians supports the general applicability of the QCM method to quantum many-body problems.

We now investigate the behaviour of $E_0^{\rm L(4)}$ against $\langle H \rangle$ and $E_0^{\rm CMX(5)}$ under a more realistic noise model incorporated into the simulation of the RVB variational circuits. These results are summarised in Figure~\ref{Fig4}. For the uniform 12-qubit Heisenberg model, optimal RVB circuit parameters were found under zero noise simulation, with the convergence of all three ground state energy estimates shown in Figure~\ref{Fig4}(a). Using the device backend noise model framework from Qiskit~\cite{Qiskit}, these circuits were then run again under a noisy simulation designed to mimic our real experiment from Figure~\ref{Fig2} on the \textit{ibmq\_guadalupe} device and the resulting energy estimates from the computed moments are shown in Figure~\ref{Fig4}(b). Here, we have included thermal relaxation, depolarisation and readout error using calibration data from \textit{ibmq\_guadalupe} at the time of running the circuits. The gate thermal relaxation is determined by the average $T_1$, $T_2$ and gate times. The resulting error rate is then calculated, and the discrepancy between it and the randomised benchmarking error rate on the device is accounted for in a depolarising channel. For comparison, the results from running these experiments on \textit{ibmq\_guadalupe} are shown in Figure~\ref{Fig4}(c).
We see that, although both moments-based estimates have better convergence to the ground state energy at zero noise than $\langle H \rangle$, only $E_0^{\rm L(4)}$ displays any noise robustness. These results confirm the content of Figure~\ref{Fig3}, and allow us to conclude that $E_0^{\rm L(4)}$ is in practice an overall superior NISQ moments-based estimate to $E_0^{\rm CMX(5)}$, which requires computation of a higher order Hamiltonian moment and shows no robustness to quantum errors.

Each of the readout and gate error parameters in the noisy simulation were all multiplied by a factor $\alpha$. Figure~\ref{Fig4}(d) shows the results from running these simulations for a range of values of $\alpha \in [0,1]$, to get an idea of how well the QCM approach would perform on NISQ devices with reduced error rates of $\alpha$ times that which is presently available. We see that QCM retains its usefulness as an error mitigation scheme even at lower error rates. With no error mitigation techniques applied, the variational estimate $\langle H \rangle$ only comes in range of $E_0^{\rm L(4)}$ when $\alpha$ is of order $10^{-2}$, suggesting that in order to achieve a similar precision as the QCM method from the raw variational calculation, the average CNOT fidelities in a NISQ device 
would need to increase to~$\sim99.99\%$.

The main comparison we have made throughout our analysis has been between the QCM energy estimate $E_0^{\rm L(4)}$ and the energy expectation value $\langle H \rangle$, but the extra quantum circuits (or shots) required to compute $E_0^{\rm L(4)}$ raise the question of whether this is a fair comparison to make. However, simply performing additional shots to compute an expectation value only reduces statistical error, not the noise arising from quantum decoherence. In our experiments on real devices (Figure~\ref{Fig1} and Figure~\ref{Fig2}), we have used sufficient shots that the effect of this statistical noise is negligible in comparison to the quantum noise. The simulations shown in Figure~\ref{Fig3} and Figure~\ref{Fig4} have confirmed that $E_0^{\rm L(4)}$ is more robust to quantum noise than $\langle H \rangle$ by dealing with density matrices directly, effectively using infinite shots to compare these quantities. For a fairer comparison, one may consider spending the additional resources required to measure the higher order moments of the Hamiltonian instead on error mitigation strategies such as zero noise extrapolation~\cite{endo2018practical,giurgica2020digital} and probabilistic error cancellation~\cite{temme2017error,bravyi2021mitigating}. Note that, unlike with QCM, the sampling overheads of these techniques generally scale exponentially with circuit depth and noise, so for the deeper circuits studied in this work, they may not be feasible. Knowing the regimes in which these other approaches outperform QCM in terms of error suppression is a nontrivial avenue of future research, as the extra resources we would allocate to such tasks for fair comparison with QCM depends on the Hamiltonian. For example, $\langle H \rangle$ for the 12 qubit Heisenberg examples in Figure~\ref{Fig2} can be computed with 3 basis measurements, rather than 990 for moments up to $\langle H^4 \rangle$, so the same resources could be used in sampling $\langle H \rangle$ hundreds of times to correct for errors. However, for Ising type models, the QCM approach does not require any further quantum computations since all of the terms in $\left\{H, H^2, H^3, H^4\right\}$ commute, and thus any error mitigation strategy would necessarily require more overhead than QCM. In any case we emphasise that QCM is not a replacement for these quantum error mitigation techniques, but rather is fully compatible with them as they can be used to improve the estimates of the individual moments.

\section{Conclusion} \label{sec4}
In this work, we have presented and analysed the error robustness of the QCM approach when applied to a variety of quantum many-body ground state energy problems. This showcases the ability of the quantity $E_0^{\rm L(4)}$ to effectively filter out noise generated in an ansatz circuit. The quantum computation of moments can supplement an otherwise ineffectual VQE output to obtain additional information about the ground state from the Hamiltonian itself. We showed this approach to be effective for a variety of cases of up to 20 qubits on real hardware (without error mitigation), representing some of the largest stable quantum energy computation results in the literature.

An important follow-up question concerns how our method compares to classical heuristics designed to solve similar problems. Directly comparing quantum and classical approaches is a nontrivial task. In particular, complexity of the task to estimate ground state energies is heavily dependent on both the form of the subject Hamiltonian and the desired tolerance~\cite{weimer2021simulation,lee2022there}. To benchmark the QCM method, we identified one such tolerance value, $E^{\rm L(4)}_{\rm HT}$, above which the ground state energy can be approximated efficiently via the classical computation of maximally mixed moments. We find that, in most cases of a well chosen ansatz on current NISQ hardware and under noisy simulation, QCM finds a ground state energy estimate within this tolerance. This evidence is in contrast to the NISQ \textit{prima facie} view that useful information cannot be extracted from a noisy quantum computation over a typical circuit depth required by VQE~\cite{brandhofer2021error}.

Future questions for both the feasibility and utility of QCM remain.
We expect that greater precision can be attained by considering higher order moments~\cite{witte} -- the current approach or one similar to the quantum power method in~\cite{seki2021quantum} may be used to achieve this, and the key questions would be whether the noise robustness persists and how the resulting computational overhead scales to larger problems.
There is also scope for improving the measurement efficiency of QCM via other approaches such as derandomised shadows~\cite{huang2021efficient} and general commutation partitioning~\cite{gokhale2020n}.
So far we have only considered ground state energy problems, however there are approaches one could take to estimate other properties of the ground state using quantum computed moments~\cite{hollenberg1994staggered}. It is of interest how well we can extract this information from imperfect trial states and, more importantly, whether the noise robustness observed in this work persists when looking at other ground state observables under these different schemes.

Another future direction of research concerns the VQE optimisation process. QCM has thus far been studied as a method for extracting accurate noise-robust energy estimates with respect to an optimal trial state obtained by varying $\langle H \rangle$ under zero noise simulation, rather than as a method applied during the optimisation procedure. At the beginning and throughout the optimisation, $E_0^{\rm L(4)}$ does not work as a variational cost function in the same way as $\langle H \rangle$ as it is not an upper bound of $E_0$, and for many problems the computational overhead could grow quite large if all moments were to be computed at each iteration of the optimiser. However, due to the noise robustness of moments-based energy estimates we have observed in this work, it is reasonable to envision that QCM may have some utility in mitigating noise-induced effects~\cite{wang} that hinder the optimisation process as well.

If used in conjunction with the QCM approach, additional error mitigation strategies~\cite{endo2018practical,giurgica2020digital,temme2017error,bravyi2021mitigating} should serve to improve the estimate of Hamiltonian moments from the quantum computer, thus giving a more accurate value for the QCM ground state energy estimate. We emphasise that distinct from these typical error mitigation techniques -- which demand an exponential increase in resources -- the QCM method requires only polynomial overhead in the number of measurements. This, balanced with the sharp decrease in effective error-per-gate shown by our results, suggests that avenues to quantum advantage in the NISQ era may be far more attainable than previously suspected.

\section{Acknowledgements}
This research was supported by the University of Melbourne through the establishment of the IBM Quantum Network Hub at the University.
HJV, MAJ, GALW, and FMC are each supported by an Australian Government Research Training Program Scholarship. CDH was supported through a Laby Foundation grant at The University of Melbourne. 
This research was supported by The University of Melbourne’s Research Computing Services and the Petascale Campus Initiative.

\bibliographystyle{quantum}

\end{document}